\documentclass[prd,nofootinbib,floats,superscriptaddress,eqsecnum,tightenlines,11pt]{revtex4}
\pdfoutput=1
\usepackage{hyperref}
\usepackage{graphicx}
\usepackage{amsmath,amssymb,amsfonts,amsthm,latexsym,stmaryrd}
\usepackage{marginnote}
\usepackage{color}
\usepackage{soul}

\newcommand{\nb}{\nonumber}
\def\beq{\begin{equation}}
\def\eeq{\end{equation}}
\newcommand{\bea}{\begin{eqnarray}}
\newcommand{\eea}{\end{eqnarray}}
\def\bfig{\begin{figure}}
\def\efig{\end{figure}}

\newcommand{\Fz}{P}
\newcommand{\Hb}{H_b}
\newcommand{\cA}{\beta}
\newcommand{\cL}{\mu}
\newcommand{\cP}{\alpha}

\newcommand{\R}{R}

\usepackage[stable]{footmisc}

\def\dkmu2{\delta K_{\mu \nu}\delta K^{\mu \nu}}
\def\pmu2{  \phi_{\mu \nu}\phi^{\mu \nu}}

\renewcommand\[{\left[}

\newcommand\ees{\end{eqnarray}}
\newcommand\bees{\begin{eqnarray}}

\newcommand\alphaB{\alpha_{\text{B}}}

\newcommand\alphaK{\alpha_{\text{K}}}
\newcommand\alphaT{\alpha_{\text{T}}}
\newcommand\alphaH{\alpha_{\text{H}}}

\def\a{\alpha}

\newcommand{\quadac}{{\rm  quad}}
\newcommand{\As}{A}

\newcommand{\aL}{\alpha_{\rm L}}

\newcommand{\bun}{\beta_1}
\newcommand{\bdeux}{\beta_2}
\newcommand{\btrois}{\beta_3}
\newcommand{\A }{\As  } 

\begin{document}

\title{Late time cosmological evolution in  DHOST models}

\author{Hamza Boumaza}
\affiliation{Laboratory of Theoretical Physics and Department of Physics,Faculty of Exact and Computer Sciences, University Mohamed Seddik Ben Yahia, BP 98, Ouled Aissa, Jijel 18000, Algeria}
\author{David  Langlois}
\affiliation{Astroparticule et Cosmologie (APC), \\
CNRS, Universit\'e de Paris, F-75013 Paris, France}
\author{Karim  Noui}
\affiliation{Institut Denis Poisson, CNRS, Universit\'e de Tours, 37200 Tours, France}
\affiliation{Astroparticule et Cosmologie (APC), \\
CNRS, Universit\'e de Paris, F-75013 Paris, France}
\date{\today}

\begin{abstract}
\noindent 
We study the late cosmological evolution, from the nonrelativistic matter dominated era to the dark energy era,  in modified gravity models described by Degenerate Higher-Order Scalar-Tensor (DHOST) theories. They represent  the most general scalar-tensor theories propagating a single scalar degree of freedom and include Horndeski and Beyond Horndeski theories. We provide the homogeneous evolution equations for  any quadratic DHOST theory, without restricting ourselves  to theories where the speed of gravitational waves coincides with that of light since the  present constraints apply to wavelengths much smaller than cosmological scales. To illustrate the potential richness  of  the cosmological background evolution in these theories, we  consider a simple family of shift-symmetric models, characterized by three parameters and compute the evolution of dark energy and of its equation of state. We also identify the regions in parameter space
where the models are perturbatively stable. 
\end{abstract}

\maketitle

\section{Introduction}
One possible explanation for the observed acceleration of the cosmological expansion is that gravity is modified on cosmological scales. Concrete realisations of this idea often rely on scalar-tensor theories, which represent the simplest extension  of general relativity since  a scalar degree of freedom is added to the usual tensor modes of general relativity. The most general family of scalar-tensor theories  that has been developed so far is that of Degenerate Higher-Order Scalar-Tensor (DHOST) theories~\cite{Langlois:2015cwa}, 
which encompass Horndeski theories~\cite{Horndeski:1974wa},   Beyond Horndeski (or GLPV) theories~\cite{Gleyzes:2014dya}  which are earlier extensions of Horndeski, as well as disformal transformations of the Einstein-Hilbert action~\cite{Zumalacarregui:2013pma}. 
In the present work, we consider the whole
family of quadratic DHOST theories, introduced in \cite{Langlois:2015cwa}  (see also \cite{Crisostomi:2016czh,Achour:2016rkg} for further details and  \cite{Langlois:2018dxi} for a review), but for simplicitly, we do not include DHOST theories with cubic terms (in second derivatives of the scalar field) which have been fully classified in   \cite{BenAchour:2016fzp}.

Most of the literature has recently concentrated on DHOST theories where the speed of gravitational waves coincides with that of light, following the  observation of a neutron star binary merger that has set an impressively stringent constraint on the difference between these two velocities \cite{TheLIGOScientific:2017qsa}.
Moreover, it has  been pointed out that subsets of DHOST theories can lead to the decay of gravitational waves, yielding a further tight constraint 
on DHOST theories~\cite{Creminelli:2018xsv,Creminelli:2019nok}.
The cosmology of DHOST theories satisfying either the first or both of the above constraints has been studied in  \cite{Crisostomi:2017pjs,Crisostomi:2018bsp,Frusciante:2018tvu,Hirano:2019nkz,Belgacem:2019pkk,Arai:2019zul}.

However, it should be stressed that the LIGO-Virgo measurements probe wavelengths of order $10^3$ km, which are many orders of magnitude smaller than cosmological scales, and an effective theory describing cosmological scales might not be adequate to describe physics on much smaller length scales, as those probed by LIGO-Virgo (see \cite{deRham:2018red} for a discussion on this point). 
In the present work, we adopt the  point of view that DHOST theories apply only to cosmological scales and  cannot be extrapolated down to  astrophysical scales within the same framework\footnote{Our motivation here is that dark energy can be described by a DHOST model. DHOST theories with a very different set of parameters could still be used to describe modified gravity in astrophysical systems, but without being able to account for dark energy because of the LIGO-Virgo constraints.}. In this perspective, all the  constraints derived from GW170817 mentioned above are not directly relevant and it thus makes sense to study models that can lead to distinct propagation velocities for light and gravitational waves on cosmological scales. 

The outline of the paper is the following. In section \ref{sec_2},
starting from the most general action for quadratic DHOST theories, we derive the Friedmann equations and the scalar field equation. These results extend those obtained recently in  \cite{Crisostomi:2017pjs} and \cite{Crisostomi:2018bsp}.
As in \cite{Crisostomi:2018bsp} we introduce an auxiliary scale factor that makes the equations manifestly second-order. 
The second part of the paper is devoted to the study of a simple subfamiliy of quadratic DHOST theories characterized by a few parameters. In section \ref{sec_3}, we write the equations of motion in the form of a dynamical system and identify the fixed points and their nature. In section \ref{sec_4}, we  turn to the linear perturbations in order to study the perturbative stability of the model. We conclude in the last section.

\section{General cosmological equations}
\label{sec_2}
In this section, we briefly recall the basic properties of DHOST theories and introduce the notations used throughout this paper. Then, we focus on the case of a homogeneous and isotropic universe and  provide the cosmological evolution equations for the whole family of quadratic DHOST theories.

\subsection{Quadratic DHOST theories}
The most general theory of quadratic DHOST theory is described by the action 
\begin{equation}
S=\int d^4x\sqrt{-g}\left(\Fz(X,\varphi)+Q(X,\varphi)\, \Box \varphi+F(X,\varphi)\,R+\sum_{i=1}^{5}A_{i}(X,\varphi)\, L_{i}\right)\label{S}
\end{equation}
where the functions $A_{i},\,F,\,Q$ and $\Fz$ depend on the scalar
field $\varphi$ and its kinetic term $X\equiv \nabla^{\mu}\varphi\nabla_{\mu}\varphi$,
$R$ is  the Ricci scalar.
The five elementary Lagrangians $L_{i}$ quadratic in second derivatives of $\varphi$  are defined by 
\begin{eqnarray}
&&L_1 \equiv \varphi_{\mu\nu} \varphi^{\mu\nu} \, , \quad
L_2 \equiv (\Box \varphi)^2 \, , \quad
L_3 \equiv \varphi^\mu \varphi_{\mu\nu} \varphi^\nu\Box \varphi \, , \quad \nonumber \\
&&L_4 \equiv  \varphi^\mu  \varphi_{\mu\nu} \varphi^{\nu\rho} \varphi_\rho \, , \quad
L_5 \equiv (\varphi^\mu \varphi_{\mu\nu} \varphi^\nu)^2 \, ,
\end{eqnarray}
where we are using the standard notations $\varphi_\mu \equiv \nabla_\mu \varphi$ and $\varphi_{\mu\nu} \equiv \nabla_\nu \nabla_\mu \varphi$ for the first and second (covariant) derivatives of $\varphi$. For the theory to be degenerate and thus propagate only one extra 
scalar degree of freedom in addition to the usual tensor modes of gravity, the functions $F$ and $A_i$ have to satisfy some conditions \cite{Langlois:2015cwa,Achour:2016rkg}
whereas $\Fz$ and $Q$ are totally free.

It has been established 
in \cite{Achour:2016rkg} that these DHOST  theories  can be classified  into three classes which are 
stable under general disformal transformations, i.e. transformations of the metric of the form
\beq
\label{disf}
g_{\mu\nu} \longrightarrow \tilde g_{\mu\nu}= C(X, \varphi) g_{\mu\nu}+ D(X,\varphi) \varphi_\mu \, \varphi_\nu\,,
\eeq
where $C$ and $D$ are arbitrary functions (provided that the metric $\tilde{g}_{\mu\nu}$ remains regular).

 The theories belonging to the first class, named class Ia in \cite{Achour:2016rkg}, can be mapped into a Horndeski form by applying a disformal transformation. 
The other two classes are not physically viable 
\cite{Langlois:2017mxy}
and will not be considered in the present work.
Theories in class Ia  are labelled by the three free functions $F,A_1$ and $A_3$ (in addition to $\Fz$ and $Q$) and 
the three remaining functions are given by the relations
\cite{Langlois:2015cwa}
\begin{eqnarray}
A_{2} & = & -A_{1} \, ,\label{deg1}
\\
A_{4} & = & \frac{1}{8\left(F+XA_{2}\right)^2}\Bigl(A_{2}A_{3}\left(16X^{2}F_{X}-12XF\right)+4A_{2}^{2}\left(16XF_{X}+3F\right)+16A_{2}\left(4XF_{X}+3F\right)F_{X}  \nonumber\\
 &  & +16XA_{2}^{3}+8A_{3}F\left(XF_{X}-F\right)-X^{2}A_{3}^{2}F+48FF_{X}^{2}\Bigr) \, \label{funA4},\\
A_{5} & = & \frac{1}{8\left(F+XA_{2}\right)^2}\Bigl(2A_{2}+XA_{3}-4F_{X}\Bigr)\Bigl(3XA_{2}A_{3}-4A_{2}F_{X}-2A_{2}^{2}+4A_{2}^{3}F\Bigr) \, ,
\label{deg3}
\end{eqnarray}
where  $F_X$ denotes the derivative of  $F(X,\varphi)$ with respect to $X$. Similarly $F_\varphi$ will denote the partial derivative of $F$ 
 with respect to $\varphi$ and the same notations will be used for all functions.
 
The above relations (\ref{deg1}-\ref{deg3}) are a direct consequence of the three degenerate conditions that
guarantee 
 only one scalar degree of freedom is present
\cite{Langlois:2015cwa,Langlois:2015skt}.
In conclusion, this means that all the DHOST theories we study here are characterized by five free functions of $X$ and $\varphi$, which are $P$, $Q$, $F$, $A_1$ and $A_3$. Notice that we have implicitly supposed the condition $F+XA_{2}\neq 0$. Theories where
$F+XA_{2}= 0$ belong to the sub-class Ib which is not physically relevant
\cite{Achour:2016rkg}.

\subsection{Homogeneous and isotropic cosmology}
We now wish to study the behaviour of these theories in a homogeneous and isotropic spacetime, endowed with the  metric
\begin{equation}
ds^{2}=-N^2(t) dt^{2}+a^2(t)\delta_{ij}dx^{j}dx^{i} \, ,\label{ds}
\end{equation}
where the lapse function $N(t)$ and the scale factor $a(t)$ depend on time only. 
As a consequence of the spacetime symmetries, the scalar field must also be homogeneous and therefore depends only on time.

Substituting the above 
metric  (\ref{ds}) into the action (\ref{S}), 
and taking into account the  degeneracy conditions (\ref{deg1}-\ref{deg3}), one finds that the corresponding homogeneous action can be written as a functional of $N(t)$, $a(t)$ and of the homogeneous scalar field $\varphi(t)$.
It reads
\begin{eqnarray}
S_{\rm hom}[N,a,\varphi] & = & \int dt\,a^{3}N\Biggl\{\Fz+Q\left(\frac{\dot{N}}{N^{3}}\dot{\varphi}-\frac{3\dot{a}}{a\,N^{2}}\dot{\varphi}-\frac{\ddot{\varphi}}{N^{2}}\right)-F_{\varphi} \frac{6\dot{a}}{a\,N^{2}}\dot{\varphi}\nonumber \\
 &  & \qquad \qquad -\frac{6f_{1}}{N^{2}}\left(\frac{\dot{a}}{a}+\frac{f_{2}}{4f_{1}}\left(\frac{\dot{N}\dot{\varphi}{}^{2}}{N^{3}}-\frac{\ddot{\varphi}\dot{\varphi}}{N^{2}}\right)\right)^{2}\Biggr\} \, , \label{Shom}
\end{eqnarray}
where we have introduced the new functions,
\begin{eqnarray}
\label{f1f2}
f_{1}  \equiv  F- X A_{1}\,,\qquad
f_{2}  \equiv  4F_{X}- 2 A_{1}+XA_{3}\,,
\end{eqnarray}
and, everywhere, the expression of  $X$ is  explicitly given by
\beq
X=-\frac{\dot{\varphi}{}^{2}}{N^{2}}\,.
\eeq

The Euler-Lagrange equations derived from the above action \eqref{Shom} lead to equations of motion that appear higher than second order. However, due to the degeneracy of the theory, these equations can be recast into a second order system.  
As done in \cite{Crisostomi:2018bsp}, 
this can be demonstrated explicitly by introducing an auxiliary scale factor $b$, defined by the relation
\beq
\label{def_b}
a\equiv \Lambda\left(X,\varphi\right)b\equiv e^{\lambda(X,\varphi)} b\,,
\eeq
where $\lambda$ satisfies the condition
\beq
\lambda_{X}=-\frac{f_{2}}{8f_{1}}\,,
\label{lambdaX}
\eeq
so that the terms quadratic in $\ddot \varphi$ in the action (\ref{Shom}) are reabsorbed in the derivatives of the new scale factor. 

It is also convenient to  use a Hubble parameter associated with this auxiliary scale factor, defined  by
\beq
\label{Hb}
H_b\equiv \frac{\dot b}{N b}=H-\lambda_X \frac{\dot X}{N} -\lambda_\varphi\frac{\dot\varphi}{N}\,, \qquad
\dot{X}= \frac{2}{N^2} \left(\frac{\dot N \dot \varphi^2}{N} - {\dot\varphi \ddot\varphi }\right) \, .
\eeq
In fact, the auxiliary variable $b$  corresponds to the scale factor of the disformally transformed metric $\tilde g_{\mu\nu}$
in (\ref{disf}) when 
the DHOST theory coincides with a Horndeski theory. The drawback of using
this ``Horndeski frame" is that matter is no longer minimally coupled, as 
it was assumed in the inital frame, which we will call  here
the ``DHOST frame''. The Horndeski and DHOST frames are, respectively, the analogs of the Einstein and Jordan frames for traditional scalar-tensor theories. 

When expressed in terms of $b$ instead of $a$, the Lagrangian $L_{\rm hom}$ in the action \eqref{Shom} becomes
\begin{eqnarray}
L_{\rm hom} & = & \Lambda^{3}b^{3}N\Biggl\{-3  \lambda_{\varphi}\frac{\dot{\varphi}{}^{2}}{N^{2}}(2 f_1 \lambda_\varphi + Q + 2 F_\varphi)+\Fz-6f_{1}H_{b}^{2} -\frac{3\dot{\varphi}}{N}\left(4\lambda_{\varphi} f_{1}+2F_{\varphi}+Q\right)H_{b} \nonumber \\
 &  & \qquad +\left(Q-6\lambda_{X}\left(2F_{\varphi}+Q\right)\frac{\dot{\varphi}2}{N^{2}}\right)\left(\frac{\dot{N}\dot{\varphi}}{N^{3}}-\frac{\ddot{\varphi}}{N^{2}}\right)\Biggr\}.
\end{eqnarray}
The coupling to matter is described by adding to $L_{\rm hom} $ a matter Lagrangian $L_m$, and the total Lagrangian is
denoted $L \equiv L_{\rm hom} + L_m$.

We get the equations of motion by writing  the Euler-Lagrange equations for $N$, $b$ and $\varphi$. 
The first two equations 
provide the generalizations of the Friedmann equations. The last equation, corresponding to the scalar field equation of motion, is obtained from an Euler-Lagrange equation of the form
\begin{eqnarray}
-\frac{d^{2}}{dt^{2}}\frac{\partial L}{\partial\ddot{\varphi}}+\frac{d}{dt}\frac{\partial L}{\partial\dot{\varphi}}-\frac{\partial L}{\partial\varphi} & = & 0.
\end{eqnarray}
Once we have derived the equations of motion, we fix the time coordinate such that  $N=1$ (and thus $\dot{N}=0$) in order to simplify the equations.

As for matter, we assume that it is 
described by a perfect fluid whose equation of state is $P=w\rho$, where $w$ is constant. The variation of the matter action $S_m$ gives  the energy-momentum tensor of the fluid, defined as usual  by
\beq
T^{\mu\nu}=\frac{2}{\sqrt{-g}}\frac{\delta S_m}{\delta g_{\mu\nu}}\,,
\qquad S_m = \int d^4x \sqrt{-g} \, L_m \, .
\eeq
As a consequence, in the DHOST frame, where matter is minimally coupled, 
the variation of the matter Lagrangian  is immediately given by
\beq
\delta L_m= - a^3 \rho_m \delta N+ 3 N a^2 P_m \delta a\,,
\eeq
where $\rho_m$ and $P_m$ are  the fluid energy density and pressure, respectively.
Using 
\beq
\frac{\delta a}{a}=-2X \lambda_X \frac{\delta N}{N}+\frac{\delta b}{b}\, 
\eeq
which follows from the definition (\ref{def_b}) of $b$, one finds that the variation of the  matter Lagrangian in the Horndeski frame is given by
\beq
\delta L_m =Nb^3 \Lambda^3 \left[-\left(\rho_m +6X\lambda_X P_m\right)\frac{\delta N}{N}+ 3 P_m \frac{\delta b}{b}\right]\,.
\eeq

Inserting the above variation of the matter  Lagrangian into the Euler-Lagrange equations, we obtain  the set of equations of motion for the cosmological dynamics. The analogs of the two Friedmann equations (with $N=1$) take the form
\begin{eqnarray}
 g_{0}+g_{1}H_{b}\dot\varphi+g_{2}H_{b}^{2} & = & \left(1-6 w \lambda_{X}\dot{\varphi}^{2}\right)  \rho_{m}  \, , \label{frw1}\\
 g_{3}+g_{4}(2\dot{H}_{b}+3H_{b}^{2})+g_{5}H_{b}\dot{\varphi}+g_{6}\ddot{\varphi}+g_{7}H_{b}\dot{\varphi} \ddot{\varphi} & = & -w \rho_{m} \,. \label{frw2}
\end{eqnarray}
where all the coefficients $g_i$ can be written explicitly in terms of the functions that appear in the Lagrangian and of $\lambda$. 
They are given in Appendix \ref{App:coeffg}.

Finally the scalar field equation can be written as
\begin{equation}
\frac{d}{dt}\left(b^{3}\Lambda^{3}J\right)+b^{3}\Lambda^{3} U=0\,,
\label{SFE}
\end{equation}
where we have defined 
\beq
J=\frac{1}{b^{3}\Lambda^{3}}\left[\frac{d}{dt}\left(\frac{\partial L_{\rm hom}}{\partial\ddot{\varphi}}\right)-\frac{\partial L_{\rm hom}}{\partial\dot{\varphi}}\right]\,, \qquad U=\frac{1}{b^{3}\Lambda^{3}}\frac{\partial L_{\rm hom}}{\partial\varphi} \,.
\eeq
The two functions $U$ and $J$ are of the form
\begin{eqnarray}
\label{UandJ}
U  =  g_{8}+g_{9}H_{b}\dot{\varphi}+g_{10}H_{b}^{2}+g_{11}\ddot{\varphi}\, , \qquad
J  =  g_{12}\dot{\varphi}+g_{13}H_{b}+g_{14}H_{b}^{2}\dot{\varphi} \,,
\end{eqnarray}
where the corresponding coefficients $g_i$ are also given explicitly in  Appendix
\ref{App:coeffg}. As usual, the equation for the scalar field is not  independent  and
can be obtained from  the two Friedmann equations. 

One recovers the equations of motion given in \cite{Crisostomi:2018bsp} when the theory is shift symmetric, i.e. when the functions in the Lagrangian are  invariant under the transformation $\varphi\rightarrow\varphi+c$ and thus depend only on $X$.
In particular, the quantity $U$ defined above vanishes and $J$ is conserved.

\section{An illustrative toy model}
\label{sec_3}
We now restrict our study to a class of models described by Lagrangians that depend on three  constant parameters only. These Lagrangians are shift symmetric  and characterized by the simple polynomial
functions 
\begin{eqnarray}
\Fz  =  \cP X, \quad Q  =  0,\quad F  =  \frac{1}{2},\quad  A_1=-A_{2}  =  -\cA X,\quad \lambda = \frac12 \cL X^2, \label{func}
\end{eqnarray}
where $\alpha,\,\cA$ and $\cL$ are arbitrary constants. Hence, from \eqref{lambdaX} and \eqref{f1f2}, we deduce
\begin{equation}
A_{3}=-2(\cA+2\cL)-8 \cA \cL X^{2}.
\end{equation}
The expressions of $A_4$ and $A_5$ are easily obtained from the degeneracy conditions \eqref{funA4} and \eqref{deg3}.
One thus gets
\beq
A_4 =  2(\beta + 2\mu- 2 \mu^2 X^2) \, , \qquad A_5 = 8\mu X (\beta + 2 \mu + 3 \beta \mu X^2)\,.
\eeq
Note that the particular choice $\mu=0$ corresponds to a subset of Horndeski theories  (in this case, the DHOST and Horndeski frames coincide, i.e. 
$\lambda=0$ and thus
$a=b$).

\subsection{Horndeski frame: dynamical system analysis}
Since we are interested in the transition between the matter and dark energy dominated eras, 
we also assume that 
matter is non-relativisitc and thus take $w=0$. 
The Friedmann-like  equations (\ref{frw1}) and (\ref{frw2}) then reduce to
\begin{eqnarray}
&&3\Hb^{2}\left(1+2\left(3\cL +5\cA\right)\text{X}^{2}+12\cL \cA \text{X}^{4}\right)-\cP\text{X}-6\cP \cL\text{X}^{3} - \rho_{m} = 0 \, ,\label{frwc1}\\
&& 3\Hb^{2}\left(1+2 \cA \text{X}^{2}\right)+2\left(1+2\cA\text{X}^{2}\right)\dot{\Hb}+\cP\text{X}-4\Hb\left(\left(3 \cL+4 \cA \right)\text{X}+6 \cA \cL \text{X}^{3}\right)\dot{\varphi}\ddot{\varphi}  =  0\,.
\label{frwc2}
\end{eqnarray}
Furthermore, the equation of motion for the scalar field  becomes
\begin{eqnarray}
&&\left[ \cP\left(1+21 \cL \text{X}^{2}+18 \cL^{2} \text{X}^{4}\right)-9 X\left(4\cA+3\cL +2\cL(11\cA+3\cL) X^2+12\cA\cL^2X^4 \right)\Hb^2\right]\ddot{\varphi} 
\nonumber
\\
&&+3\left[\cP(1+3\cL X^2)-X (4\cA+3\cL+6\cA\cL X^2)(3\Hb^2+2\dot \Hb) \right]\Hb\, \dot{\varphi}=0\,.
\label{scalareq}
\end{eqnarray}

\medskip

To study these cosmological equations, it is convenient to rewrite them  as a dynamical system (and analyse the fixed points and their stability) with the new variables
\begin{eqnarray}
\label{defofparam}
x_{1} \equiv \cP\frac{X}{\Hb^{2}},\, \quad x_{2} \equiv \cA X^{2},\, \quad s \equiv \frac{\cL}{\cA},\,\quad \epsilon_{h} \equiv \frac{\dot{\Hb}}{\Hb^{2}},\,\quad \epsilon_{\varphi} \equiv \frac{\ddot{\varphi}}{\Hb \dot{\varphi}} \, ,
\end{eqnarray}
following similar treatments  for  dark energy models (see e.g. 
\cite{DeFelice:2010pv,DeFelice:2011bh,Boumaza:2019rpt} in the context of Galileons and  \cite{Bahamonde:2017ize} for a recent review). These variables are not independent and one can easily see that
\begin{eqnarray}
\frac{dx_{1}}{d\ln b} & =& 2\left(\epsilon_{\varphi}-\epsilon_{h}\right)x_{1} \, , \label{x1'}\\
\frac{dx_{2}}{d\ln b}  &=& 4\epsilon_{\varphi}x_{2}\label{x2'}\,,
\end{eqnarray}
where $\ln b$ plays the role of time. 
The two equations above
 can also be formulated in terms of the time $\ln a$ instead of $\ln b$ by using the relation between the two Hubble constants,
\beq
\label{H_Hb}
H=(1+2 s x_{2}\epsilon_{\varphi})H_{b}\,,
\eeq
which follows from \eqref{Hb}. The two previous relations (\ref{x1'}) and (\ref{x2'}) then become
\begin{align}
x_{1}' & =\frac{2(\epsilon_{\varphi}-\epsilon_{h}) x_{1} }{1+2s \epsilon_{\varphi} x_{2}}\, , \label{x1'a}\\
x_{2}' & =\frac{4\, \epsilon_{\varphi}\, x_{2} }{1+2s\epsilon_{\varphi} x_{2}}\, , \label{x2'a}
\end{align}
where a prime denotes a  derivative with respect to $N\equiv \ln a$.

\medskip

So far, we have not yet used the equations of motion, namely  the Friedmann-like equations, Eqs.~\eqref{frwc1} and \eqref{frwc2}, and the scalar equation \eqref{scalareq}. They can be reformulated, respectively,  as
\begin{eqnarray}
&&\frac{x_{1}}{3}-12sx_{2}^{2}-6sx_{2}+2sx_{1}x_{2}-10x_{2}+\Omega_{m}  =1 \, , \label{eq:f1}\\
&&{2}(1+2x_{2})\epsilon_{h}+{4}x_{2}(4+3s+6sx_{2})\epsilon_{\varphi}+(3+x_{1}+6x_{2})  =0 \, , \label{eq:f2}\\
&&-\frac{2}{3} \left(x_1 \left(18 s^2 x_2^2+21 s x_2+1\right)-9 x_2 \left(6 s^2 x_2 \left(2 x_2+1\right)+s \left(22 x_2+3\right)+4\right)\right) \epsilon _{\varphi }\nonumber\\
&&+4 x_2 \epsilon _h \left(s \left(6 x_2+3\right)+4\right)+36 s x_2^2+18 s x_2-6 s x_1 x_2+24 x_2-2 x_1 =0\,.\label{eq:scalar}
\end{eqnarray}
These equations can be seen as constraints for the dynamical system
(\ref{x1'a}-\ref{x2'a}).  
The first constraint, Eq.~(\ref{eq:f1}), involves the matter density, whereas the last two equations (\ref{eq:f2}) and (\ref{eq:scalar}) can be used to determine  $\epsilon_{\varphi}$
and $\epsilon_{h}$ in terms of $x_1$ and  $x_2$. After a straightforward calculation, one gets
\begin{align}
\epsilon_{h} & =\left[{3}x_{1}\left(72s^{2}x_{2}^{3}+12s(3s+8)x_{2}^{2}+26x_{2}-1\right)-x_{1}^{2}\left(18s^{2}x_{2}^{2}+21sx_{2}+1\right)
\right.
\nonumber \\
 & 
 \left.
 -{9}x_{2}\left(18s^{2}x_{2}\left(2x_{2}+1\right){}^{2}+3s\left(20x_{2}^{2}+4x_{2}-3\right)+40x_{2}-12\right)\right]/\Delta
  \, , \label{epsh} \\
\epsilon_{\varphi} & =-{6}x_{1}\left(12sx_{2}^{2}+6(s+1)x_{2}+1\right)/\Delta  \, , \label{epsphi}
\end{align}
where the 
common denominator $\Delta$ is given by
\begin{eqnarray}
\Delta  &\equiv & 6x_{2}\left(18x_{2}\left(2sx_{2}+s\right){}^{2}+3s\left(20x_{2}^{2}+4x_{2}-3\right)+40x_{2}-12\right) \nonumber \\
&+& 2x_{1}\left(2x_{2}+1\right)\left(18s^{2}x_{2}^{2}+21sx_{2}+1\right) \, .
\end{eqnarray}
Hence, the equations of motion are now given in the form \eqref{x1'a} and \eqref{x2'a} with $\epsilon_h$ and $\epsilon_\varphi$
given by the equations \eqref{epsh} and \eqref{epsphi}.

The critical points are found by solving the equations 
\eqref{x1'a} and \eqref{x2'a} for $x_{1}'=0$ and $x_{2}'=0$. 
The number and stability properties of these fixed points are 
summarized in Table~\ref{table1}. We see that there are at most two stable fixed points corresponding
to a de Sitter solution. To find the conditions on the parameters of the theory
for these fixed points to exist,
we have to study the signs of $x_1$ and $x_2$ at the fixed points. It is immediate to show that
\begin{itemize}
\item at the point $C$: $x_1 <0$ and $x_2<0$ for all values of $s$;
\item at the point $D$: $x_1>0$ and $x_2<0$ for $s>0$ whereas $x_1<0$ and $x_2>0$ for $s<0$.
\end{itemize}
From the definition of $x_1$ and $x_2$ \eqref{defofparam}, we see immediately that  $\alpha x_1<0$ (because $X<0$) and $\beta x_2>0$. As a consequence, we deduce that the fixed point $C$ exists only if $\alpha>0$ and $\beta<0$ whereas the fixed point $D$
exists only if $\mu<0$.

Notice that, 
in the limit  $s=0$, i.e.  $\mu=0$, corresponding to  a DHOST theory that belongs to the Horndeski subclass, the dynamical system admits  a single fixed point given by  the limit $s\rightarrow 0$ of $C$,
\beq
x_{1}=-2\,, \qquad x_{2}=-\frac16\,,
\eeq
whereas the limit of the fixed point D is ill-defined.

\begin{table}[h]
\begin{tabular}{|c|c|c|c|}
\hline 
points & $x_{1}$ & $x_{2}$ & Eigenvalues\tabularnewline
\hline 
\hline 
A & $0$ & $x_{2}$ & $(0,3)$ (unstable)\tabularnewline
\hline 
B & $3$ & $0$ & $(-12,3)$ (Saddle)\tabularnewline
\hline 
C & ${(3-3s-\sqrt{3(3s^{2}+2s+3)})}/{2s}$ & ${(-3-3s+\sqrt{3(3s^{2}+2s+3)})}/{12s}$ & $(-3,-3)$(Stable)\tabularnewline
\hline 
D & ${(3-3s+\sqrt{3(3s^{2}+2s+3)}}/{2s}$ & $-{(3+3s+\sqrt{3(3s^{2}+2s+3)})}/{12s}$ & $(-3,-3)$(Stable)\tabularnewline
\hline 
\end{tabular}
\caption{Fixed points of the dynamical system with the eigenvalues of the corresponding Hessian matrix. Only the last
two points are stable.}
\label{table1}
\end{table}

\vskip 1cm

\subsection{DHOST frame: Effective Friedmann equations}

In the frame where matter is minimally coupled, it is always possible to write effectively the Friedmann equations in the usual form,
\beq
\label{Friedmann_eff}
3H^2= \rho_m+\rho_{\rm DE}\,,\qquad 2\dot H+3H^2=P_m +P_{\rm DE}\,,
\eeq
where all new terms are ``hidden"
 in the effective dark energy density and pressure, denoted $\rho_{\rm DE}$ and $P_{\rm DE}$ respectively. 
Hence, one can also define an equation of state parameter $w_{\rm DE}$ for dark energy as usual by the ratio
\beq
w_{\rm DE}\equiv \frac{P_{\rm DE}}{\rho_{\rm DE}}\,.
\eeq
Moreover,  one can define a global effective equation of state parameter as
\beq
\label{w_eff}
w_{\rm eff}\equiv \frac{P_m +P_{\rm DE}}{ \rho_m+\rho_{\rm DE}}=-1-\frac{2}{3}\frac{\dot H}{H^2}\,.
\eeq
For the models we are considering here, this parameter can be expressed in terms of the variables introduced earlier and reads
\begin{equation}
w_{\rm eff} = -1-\frac{2}{3}\frac{H'}{H}=-1-\frac{2\epsilon_{h}+4s (\epsilon_{\varphi}x_{2}'+\epsilon_{\varphi}'x_{2})}{3(1+2s x_{2}\epsilon_{\varphi})} \, .
\end{equation}
Using the fact that $P_m=0$ for non-relativistic matter, we can write, from (\ref{Friedmann_eff}) and (\ref{w_eff}), a relation
between $w_{\rm DE}$ and $w_{\rm eff}$ given by
\beq
w_{\rm DE}=\frac{w_{\rm eff}}{\Omega_{\rm DE}}\,, \qquad \Omega_{\rm DE} \equiv \frac{\rho_{\rm DE}}{3H^2}\,.
\eeq

The dynamical equations (\ref{x1'}) and (\ref{x2'}) can be solved numerically and the right amount of nonrelativistic matter today, i.e. $\Omega_{m}=\rho_m/(3H^2)\approx 0.3$, can be reached by tuning the  initial conditions for $x_1$ and $x_2$. We choose our initial conditions  deep in the matter dominated era, i.e. when $\Omega_m\simeq 1$. According to the constraint (\ref{eq:f1}), taking 
$|x_1| \ll 1$ and $|x_2| \ll 1$ initially guarantees that we are deep in the matter dominated era. Moreover, in order to observe a relatively rapid transition from the matter era to de Sitter era, we take initial conditions such that  
\bea
\label{initialcondition}
|x_1| \, \ll \, |x_2| \, \ll \, 1\, .
\eea
Indeed,  in this regime,  the dynamical
system reduces to 
 \begin{eqnarray}
x_1'  \approx  3x_1 \, , \qquad
x_2'  \approx  {4 x_1}/{(9 s+12)} \, ,
\end{eqnarray}
which shows that the system moves quickly away from the region where $x_1$ and $x_2$ are very small (this is not the case if we take $|x_2| \ll |x_1| \ll 1$ instead). The initial dark energy parameter $\Omega_{\rm DE}$ is then approximated, according to  (\ref{eq:f1}), by 
\beq
\label{Omega_DE_approx}
\Omega_{\rm DE}\approx -2(3s+5)x_2\,.
\eeq
If we choose $x_2>0$ initially, 
in order  to get $c_T<1$ (as we will see in Eq.~\eqref{PlanckandcT} of the next section), 
then the parameter $s$ must satisfy $s<-5/3$. 

\begin{figure}
\includegraphics[width=6cm]{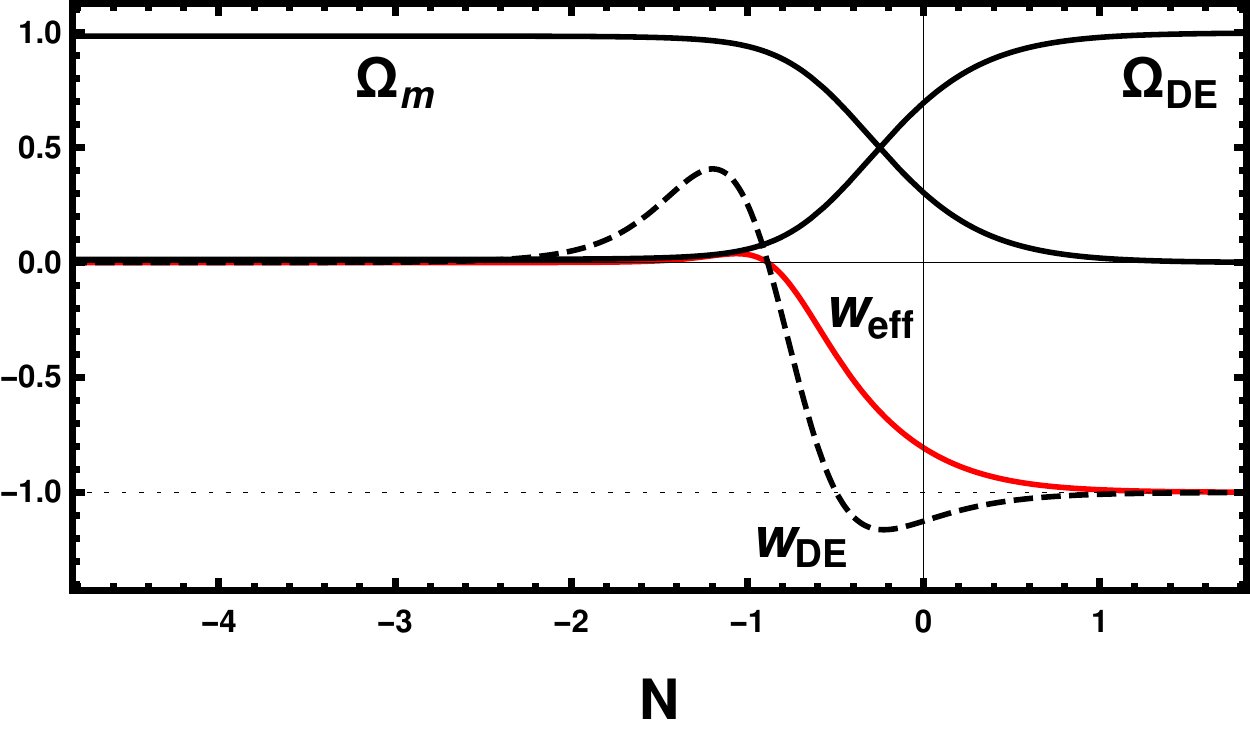}\hspace{1cm}\includegraphics[width=6cm]{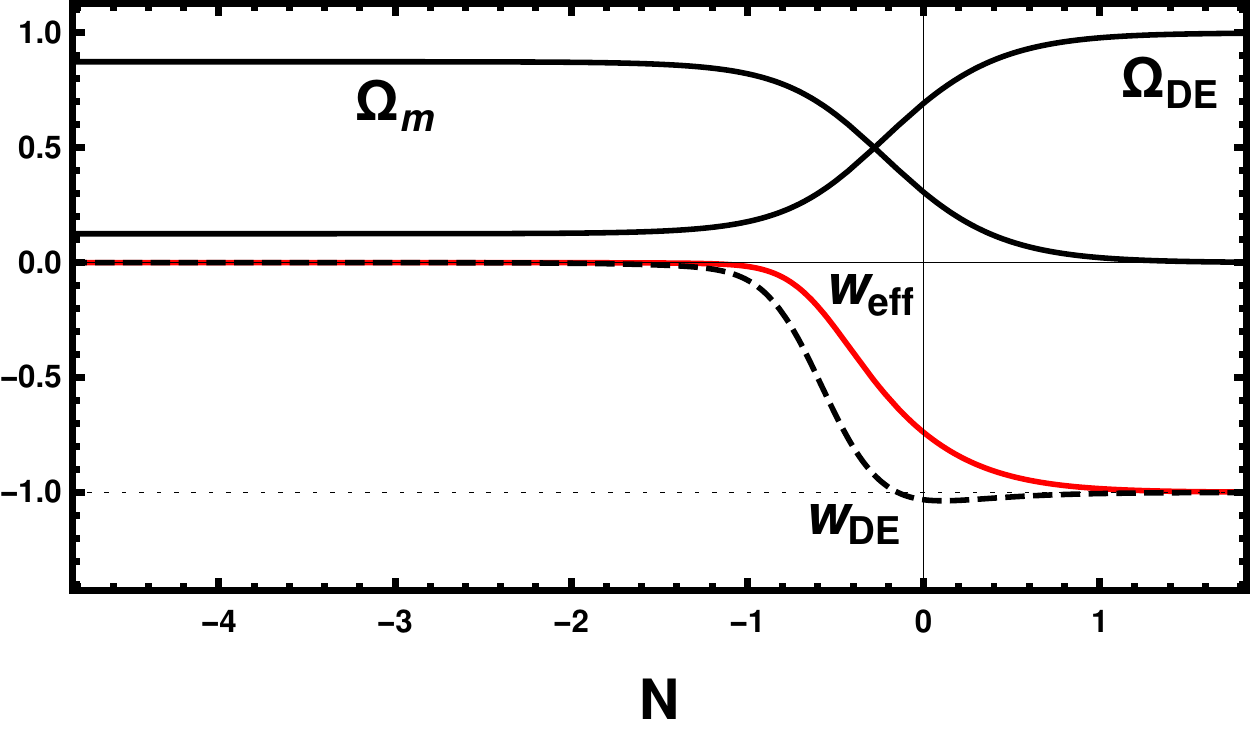}
\caption{Evolution of the parameters $w_{\rm DE}$, $w_{\rm eff}$, $\Omega_{\rm DE}$,  and $\Omega_{m}$ as functions of $N= \ln(a)$ 
for various choices for the parameter $s$ and the initial conditions. 
Left: $s=-4$ and initial conditions $x_{1}^{(i)}=-3.50\times 10^{-7}$ and $x_{2}^{(i)}= 10^{-3}$. 
Right: $s=-10$ and  initial conditions $x_{1}^{(i)}=-7.00\times 10^{-7}$ and $x_{2}^{(i)}=2.50\times10^{-3}$.}
\label{plotw}
\end{figure}

\begin{figure}
\includegraphics[width=6cm]{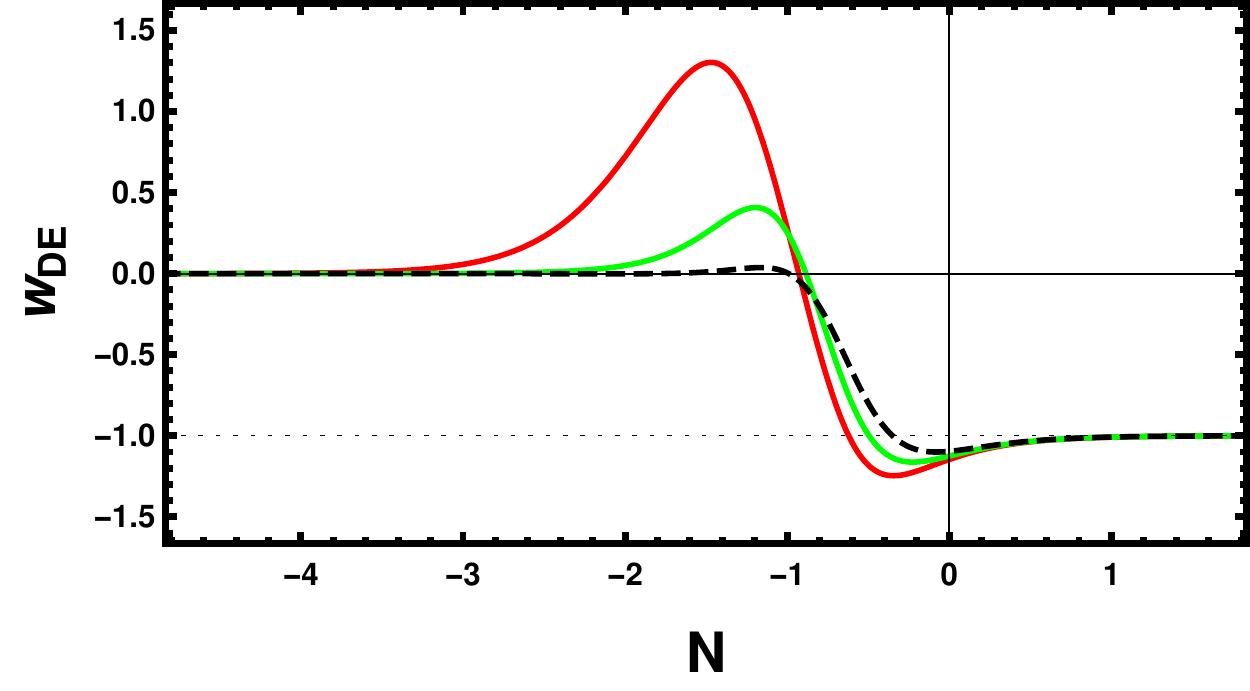}\hspace{1cm}\includegraphics[width=6cm]{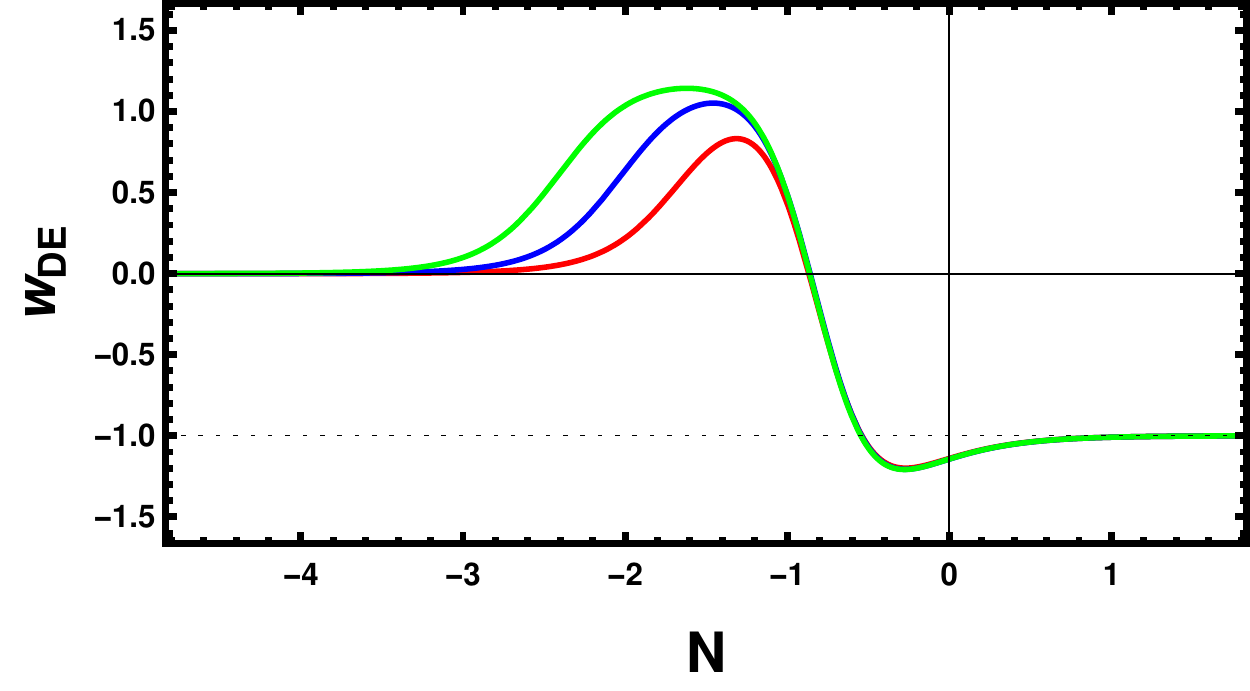}
\caption{Evolution of the dark energy ratio $w_{\rm DE}=P_{\rm DE}/\rho_{\rm DE}$ as a function of $N= \ln(a)$ 
for various choices of $s$ and initial conditions (set at $N=-5$). 
Left: $x_{2}^{(i)}=10^{-3}$;  $(x_{1}^{(i)},s)=(-2.50\times 10^{-7},-2)$ for the green curve, 
$(x_{1}^{(i)},s)=(-3.50\times 10^{-7},-4)$  for the red one and $(x_{1}^{(i)},s)=(-4.85\times 10^{-7},-10)$ for the black dashed one. 
Right: $s=-4$; $(x_{1}^{(i)},x_{2}^{(i)})=(-1.15\times 10^{-7},10^{-4})$ for the red curve, 
$(x_{1}^{(i)},x_{2}^{(i)})=(-3.60\times 10^{-8},10^{-5})$ for the blue one and $(x_{1}^{(i)},x_{2}^{(i)})=(-1.15\times 10^{-8},10^{-6})$ for the green one.}
\label{plotw22}
\end{figure}

We have plotted some illustrative examples of numerical results  in  (Fig. \ref{plotw}) and (Fig. \ref{plotw22}).
The first figure  shows the evolution of the cosmological parameters $\Omega_m$, $\Omega_{\rm{DE}}$, $w_{\rm{DE}}$
and $w_{\rm{eff}}$. We observe a cosmological transition from the matter era to the dark energy era. We also observe that the dark energy behaves like  pressureless matter  deep in the matter dominated era and  like a cosmological constant with $w_{\rm{DE}}\approx -1$ at very late times, with a transition going through an intermediate regime where $w_{\rm{DE}}$ can even reach some significant positive values.

\section{Perturbative linear stability}
\label{sec_4}

In this section, we  study  the linear  stability of the 
models studied in the previous section. 
For that purpose, we work in the framework of the Effective Theory  of Dark Energy 
developed  in  {\cite{Gubitosi:2012hu,Gleyzes:2013ooa,Gleyzes:2014rba}
and extended 
to DHOST theories in \cite{Langlois:2017mxy}. This effective approach relies on the ADM formulation where the
metric is parametrized by the lapse function $N$, the shift vector $N^i$ and the spatial metric $h_{ij}$ as follows,
\bea
ds^2 \; = \; -N^2 dt^2 + h_{ij} (dx^i + N^i dt)(dx^j + N^j dt) \, .
\eea
In the ADM framework, the ``velocity'' of the spatial metric is encoded in the extrinsic curvature tensor $K_{ij}$ defined by
\bea
K_{ij} \; \equiv \; \frac{1}{2N} \left( \dot{h}_{ij} - D_i N_j - D_j N_i \right) \, ,
\eea
where $D_i$ denotes the spatial covariant derivative associated to $h_{ij}$. The DHOST action can be reformulated in terms of the ADM variables
and the dynamics of the linear perturbations about an FLRW background is governed by the expansion of this action at  quadratic order
in the variables $\delta N$, $\delta K_{ij}$ and $\delta h_{ij}$. After a long 
but straightforward 
calculation, one finds that the quadratic action for the perturbations
is given by  \cite{Langlois:2017mxy}
 \bea
\begin{split}
\label{SBAction0}
& S^\quadac = \int d^3x \,  dt \,  a^3  \frac{M^2}2\bigg\{ \delta K_{ij }\delta K^{ij}- \left(1+\frac23\aL\right)\delta K^2  +(1+\alphaT) \bigg( \R \frac{\delta \sqrt{h}}{a^3} + \delta_2 R \bigg)\\
&  + H^2\alphaK \delta N^2+4 H \alphaB \delta K \delta N+ ({1+\alphaH}) \R  \delta N   +  4 \bun  \delta K  {\delta \dot N }   + \bdeux  {\delta \dot N}^2 +  \frac{\btrois}{a^2}(\partial_i \delta N )^2   
\bigg\} \; ,
\end{split}
\eea
where $\delta^2R$ stands for the second order term in the perturbative expansion of the Ricci scalar $R$ and  $h$ is the determinant of the spatial metric. The
coefficients  $M$, $\aL$, $\alphaT$, $\alphaK$, $\alphaB$, $\alphaH$, $ \bun$, $\bdeux$ and  $\btrois$, which 
fully characterize the quadratic action, are functions of time  as 
they depend on the background. They can be expressed explicitly  in terms of the functions entering the DHOST action  \eqref{S}, as recalled in the Appendix \ref{EFTparam}.

After integrating out the gauge degrees of freedom, and ignoring the coupling to matter for the moment,  it has been shown in 
\cite{Langlois:2017mxy} that the quadratic action reduces to the sum of an action for 
the curvature perturbation $\zeta$, representing the scalar mode,
\bea
S_{\rm quad}[\zeta] \; = \; \int d^3x \, dt \, a^3 \frac{M^2}{2} \left[ A_\zeta \, \dot \zeta^2 
- B_\zeta \, \frac{(\partial_i \zeta)^2}{a^2}   \right] \, ,
\eea
and an action for the tensor modes $\gamma_{ij}$,
\bea
S_{\rm quad}[\gamma_{ij}] \; = \; \int d^3x \, dt \, a^3 \frac{M^2}{8} \left[ \dot \gamma_{ij}^2 - \frac{c_T^2}{a^2} (\partial_k \gamma_{ij})^2\right] \,.
\eea
The coefficients  $A_\zeta $ and $B_\zeta $ that appear in the scalar action
are given by
\bea
A_\zeta  &=&  \frac{1}{(1+\alpha_B -\dot \beta_1/H)^2} \left[\alpha_K + 6 \alpha_B^2 - \frac{6}{a^3 H^2 M^2} \frac{d}{dt}\left(a^3 H\, M^2 \alpha_B \beta_1\right)\right]  \,, \label{Azeta} \\ [2ex]
B_\zeta  &=& 
-2(1+\alpha_T) + \frac{2}{a M^2} \frac{d}{dt} \left[\frac{a M^2 \left(1+\alpha_H + \beta_1(1+\alpha_T) \right)}{H(1+\alpha_B) - \dot \beta_1}\right] \,, \label{Bzeta}
\eea
while the speed of gravitational waves $c_T$, 
which appears in the tensor action,  is given by $c_T^2=1 + \alpha_T$. Therefore the stability conditions for the linear perturbations
are simply given by
\bea
M^2>0 \, , \qquad
A_\zeta > 0 \, , \qquad
B_\zeta > 0 \, , \qquad
c_T^2 >0 \, .  
\eea
The expressions of these coefficients in terms of the dynamical variables \eqref{defofparam} are given in Appendix \ref{EFTparam}.
As we will see, they will be useful for the numerical analysis of the linear stability of the model.

\medskip

In the presence of 
matter, these stability conditions (for the scalar mode) are 
modified. They have been derived explicitly  in  \cite{Langlois:2017mxy}
for the simple case where matter is described by a scalar field $\sigma$ whose dynamics is governed by a k-essence type action,
\bea
S_m \; = \; \int d^4x \, \sqrt{-g} \, K(Y) \, , \qquad Y \equiv g^{\mu\nu} \sigma_\mu \sigma_\nu \, ,
\eea 
which is added to the DHOST action. 
The link  with a perfect fluid description of matter, with energy density $\rho_m$, pressure $P_m$ and sound speed $c_m$ is given by the expressions
\beq
\rho_m=2 Y K_Y-K\,,\qquad  P_m=K\,, \qquad
c_m^2 \; = \; \frac{K_Y}{ K_Y +2 Y K_{YY}} \, ,
\eeq
where all terms are evaluated on a background solution.

It has been shown in  \cite{Langlois:2017mxy} that the conditions for the stability of scalar linear perturbations are modified
and  more involved than the case without matter. Indeed, in addition to $\zeta$, there is an extra scalar degree of freedom
that we denote $\delta \sigma$, and the dynamics of the two modes are entangled. The quadratic action for these two scalar perturbations
takes the form \cite{Crisostomi:2018bsp}
\bea
S_{\rm quad}=\int d^3x \, dt \, a^3 \frac{M^2}{2} \left[  \dot V^T \, {\bf K} \, \dot V - \frac{1}{a^2} \partial_i V^T \, {\bf G} \, \partial^i V + \dots \right] \,,
\eea
where the vector $V^T=(\zeta,H \frac{\delta\sigma}{\dot\sigma})$  contains the two scalar degrees of freedom and the dots stand for the terms with fewer than two (space or time) derivatives, which are not relevant for the stability discussion. The kinetic and gradient matrices read (see \cite{Crisostomi:2018bsp} for details)
\bea  {\bf K} = \left(
\begin{matrix}
    A_\zeta + \frac{\rho_m(1+w_m)}{M^2 c_m^2 \left( H(1+\alpha_B) - \dot \beta_1\right)^2}    \qquad   &  \frac{ \rho_m (1+w_m) \left( 3 c_m^2 \beta_1 -1 \right)}{M^2 c_m^2 H\left( H(1+\alpha_B) - \dot \beta_1\right)} \\[4ex]
    \frac{ \rho_m (1+w_m) \left( 3 c_m^2 \beta_1 -1 \right)}{M^2 c_m^2 H \left( H(1+\alpha_B) - \dot \beta_1\right)}       & \frac{\rho_m(1+w_m)}{M^2 c_m^2 H^2}
\end{matrix} \right) \,,  \\[4ex]
{\bf G} = \left(
\begin{matrix}
   B_\zeta    &  -\frac{ \rho_m (1+w_m) \left( 1 + \alpha_H + (1+ \alpha_T) \beta_1\right)}{M^2  H\left( H(1+\alpha_B) - \dot \beta_1\right)} \\[4ex]
    -\frac{ \rho_m (1+w_m) \left( 1 + \alpha_H + (1+ \alpha_T) \beta_1\right)}{M^2 H \left( H(1+\alpha_B) - \dot \beta_1\right)}     &  \frac{\rho_m(1+w_m)}{M^2H^2}
\end{matrix} \right)  \,. \\ \nb
\eea
 In order to avoid ghost and gradient instabilities, both matrices $\bf K$ and $\bf G$ must be positive definite. 
When matter satisfies $c_m \ll 1$ and $w_m \ll 1$, one can expand the expressions of the eigenvalues of $\bf K$ and $\bf G$ with respect to $c_m$ and $w_m$ and one obtains, at leading order,
\begin{equation}
\begin{split}
\lambda_{K_{1}}  & =  \frac{A_{\zeta}M^{2}H^{2}(1+\alpha_{B}-\beta_{1}')^{2}+6\rho_{m}\beta_{1}}{M^{2}H^{2}(1+(1+\alpha_{B}-\beta_{1}')^{2})}\,, \quad \;
\lambda_{K_{2}} = \frac{\rho_{m}}{c_{m}^{2}H^{2}M^{2}}\left[\frac{1}{(1+\alpha_{B}-\beta_{1}')^{2}}+1\right],\\
\lambda_{G_{\pm}} & = 
 \frac{B_{\zeta}}{2}+\frac{1}{2M^{2}}\left[\frac{\rho_{m}}{H^{2}}\pm\sqrt{\frac{4\rho_{m}^{2}(1+\alpha_{H}+(1+\alpha_{T})\beta_{1})^{2}}{H^{4}(1+\alpha_{B}-\beta_{1}')^{2}}+(\frac{\rho_{m}}{H^{2}}
 -M^{2}B_{\zeta})^{2}}\, \right]\,,
\end{split}
\end{equation}
where $\lambda_{K_{1,2}}$ and $\lambda_{G_\pm}$ are the eigenvalues of $\bf K$ and $\bf G$ respectively.
One thus finds that $\lambda_{K_{2}}$ is always positive 
while the sign of the three other eigenvalues 
depends on the specific background solution. 

All eigenvalues can be expressed in terms of $x_1$, $x_2$ and $\Omega_m$. Moreover, the coefficients $M^2$ and $c_{\rm T}^2$ which appear in the tensor action are given explicitly by
\beq
\label{PlanckandcT}
M^2=1+2 x_2\,, \qquad c_{\rm T}^2=\frac{1}{1+2 x_2}\,.
\eeq
Deep in the matter dominated era when $|x_1| \ll|x_2| \ll 1$,  the leading order behaviour of the eigenvalues $\lambda_{K_{1}}$ and  $\lambda_{G_{\pm}}$ is given by
\bea
\lambda_{K_{1}} \approx -9 \left(3s+4\right)x_2 \, , \qquad \lambda_{G_{+}} \approx 6-17 \left(s+1\right)x_2\, ,  \qquad 
\lambda_{G_{-}}\approx -5 \left(s+1\right)x_2 \,.
\label{eigeninmatter}
\eea
and they are all positive when we take $0<x_2\ll 1$ and $s<-5/3$, as discussed below (\ref{Omega_DE_approx}).

In Fig. (\ref{plotw2}),  we plot the time evolution of the eigenvalues, as well as  $c_T^2$. We have chosen parameters and initial conditions such 
that all eigenvalues remain positive and $c_T^2 <1$. 
With theses choices, we see that the tensor and scalar perturbations remain stable from the matter era to the
de Sitter era.

\begin{figure}[h]
\includegraphics[width=6cm]{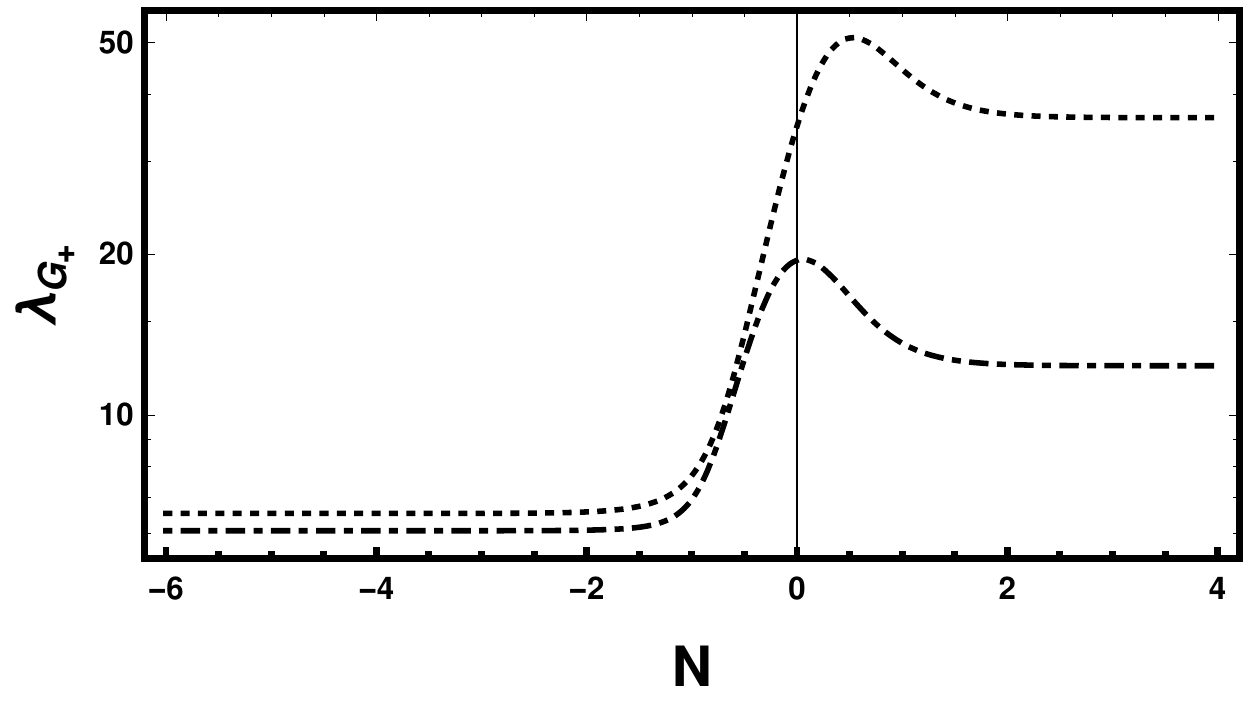} \hspace{1cm}
\includegraphics[width=6cm]{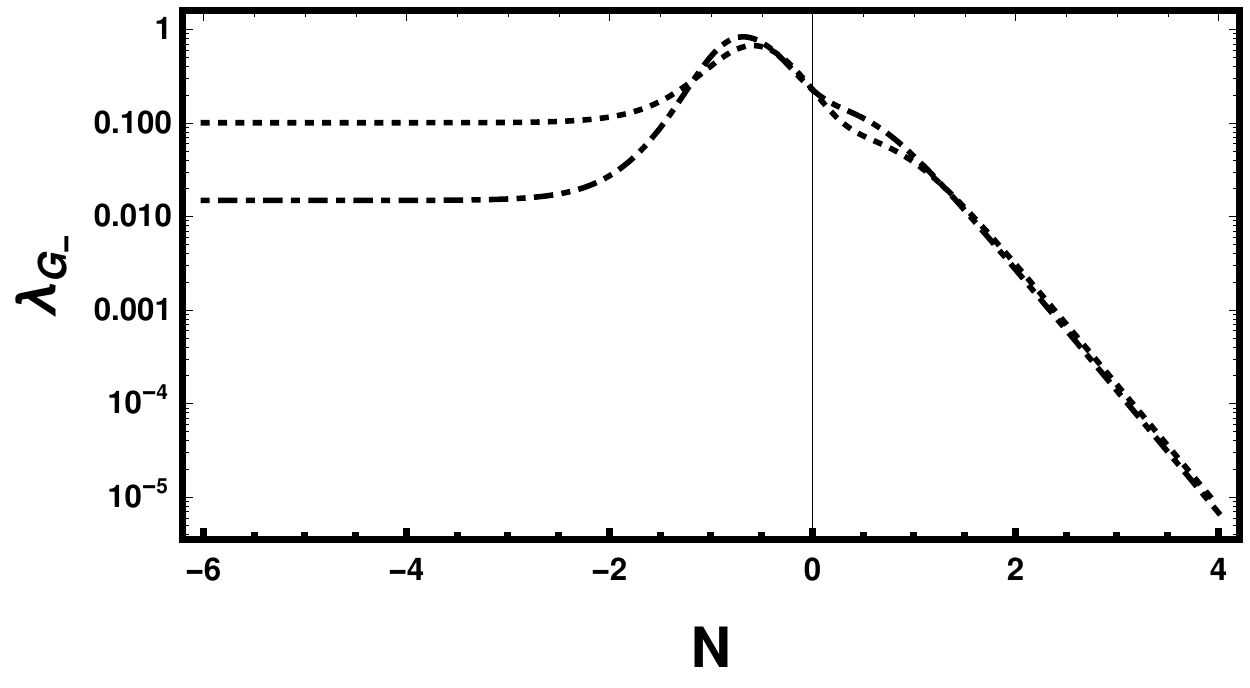} 
\includegraphics[width=6cm]{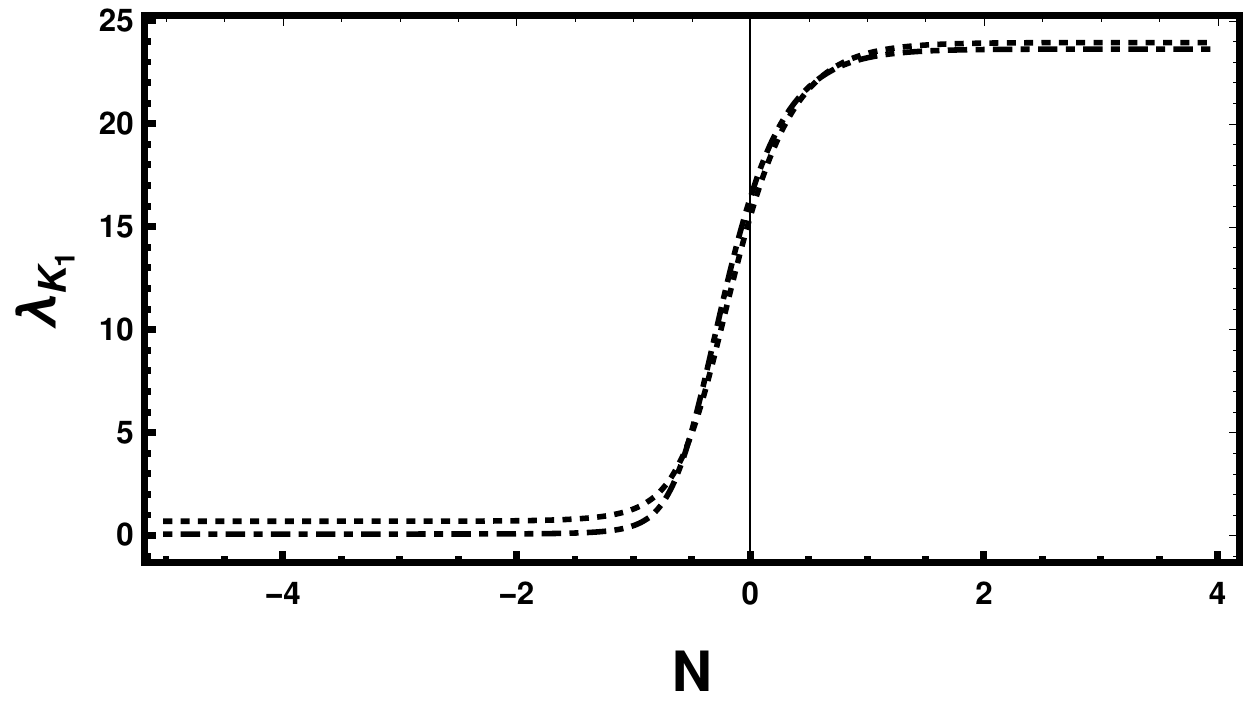} \hspace{1cm} \includegraphics[width=6cm]{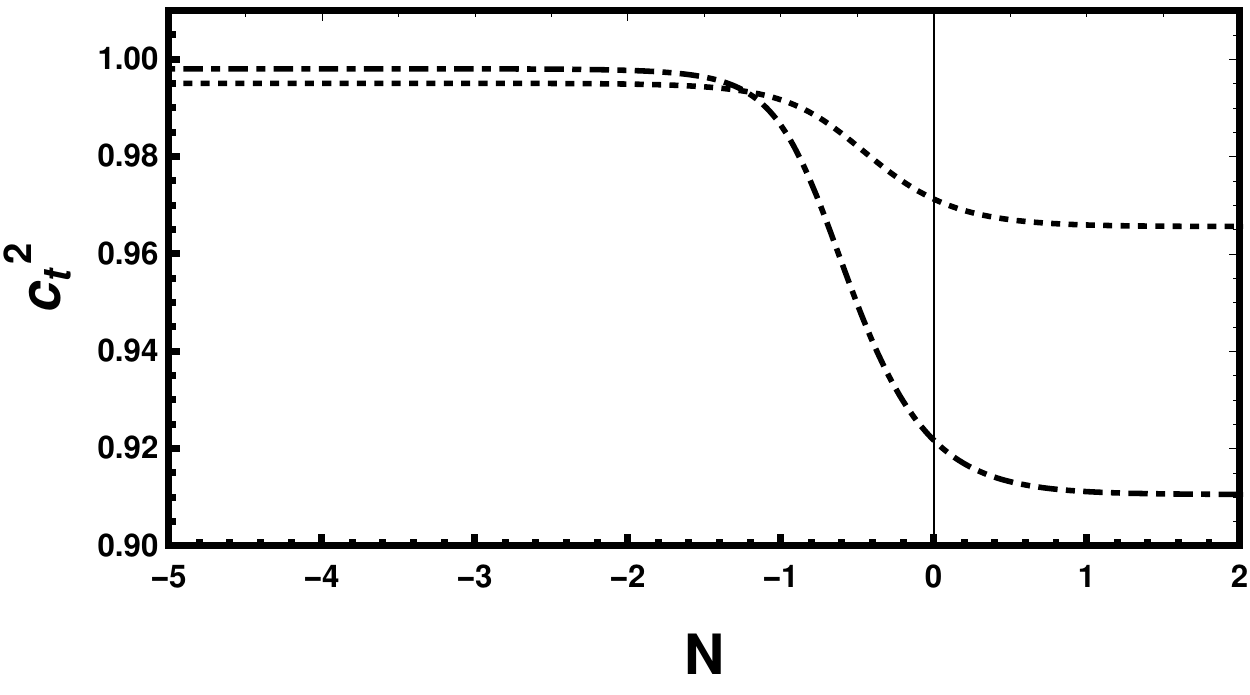}
\caption{Evolution of the eigenvalues  $\lambda_{G_\pm}$ (top), $\lambda_{K_1}$ (bottom left) and $c_T^2$ (at the bottom right) viewed as functions of $N=\ln(a)$.
{Dotted curves: $s=-4$ and initial conditions $x_{1}^{(i)}=-3.50\times 10^{-7}$ and $x_{2}^{(i)}= 10^{-3}$. 
Dot-Dashed curves: $s=-10$ and  initial conditions $x_{1}^{(i)}=-7.00\times 10^{-7}$ and $x_{2}^{(i)}=2.50\times10^{-3}$.}\label{plotw2}}
\end{figure}

\section{Conclusions}
In this paper, we have studied the cosmology of DHOST theories. We have considered the most general action for quadratic DHOST theories
and derived the general equations of motion in an isotropic and homogeneous background in the presence of  a perfect fluid. We have presented  these
equations in full generality, without restricting ourselves to  shift-symmetric Lagrangians in the first part. Then, we have considered  a particular family of shift-symmetric  DHOST models
characterized by three parameters only. We have performed a dynamical system analysis and obtained  the conditions for our models to admit self-accelerating solutions at late time. Then, we have examined   the  linear stability of both  tensor and  scalar modes, in the presence of  pressureless matter, and found that the models studied here 
are stable in some region of the parameters space. 

With the advent of stage IV cosmological probes  (LSST, Euclid), the sharp increase in the amount of data will enable us to test gravitational laws on cosmological scales. In order to analyse such a trove of data, it will be  very useful to rely on a parametrized set of models that can quantify, in a flexible way, deviations from general relativity. DHOST theories, which  describe the most general and simplest scalar-tensor theories (the simplest in the sense that they propagate a single additional degree of freedom), are natural candidates to serve as 
benchmark models for the analysis of future data.

\acknowledgements
We thank Marco Crisostomi for instructive discussions and for providing the full expressions for the parameters $\alpha_K$ and $\alpha_B$ in terms of the functions of the general Lagrangian. H. B. would like to thank APC for their hospitality during his two stays when this project was initiated and  then continued.

\appendix

\section{Coefficients in the cosmological equations}
\label{App:coeffg}
The coefficients entering in the Friedmann equations \eqref{frw1} and \eqref{frw2}  are given by
\begin{eqnarray*}
g_{0} & = & \Fz+X\left[-6\left(F\lambda_{\varphi}^{2}+F_{\varphi}\lambda_{\varphi}+\Fz\lambda_{X}\right)-2P_{X}+Q_{\varphi}\right]
\\
 &  & -6X{}^{2}\left[\lambda_{\varphi}^2( 2F_{X}+6F \lambda_{X}-3A_{1})+\lambda_{\varphi}(2F_{\text{\ensuremath{\varphi X}}} +4F\lambda_{X\ensuremath{\varphi}}+Q_{X})-2F_{\varphi\varphi}\lambda_{X}-Q_{\varphi}\lambda_{X}\right]\\
 & & +12\lambda_{\varphi}X{}^{3}\left[\lambda_{\varphi}\left(3A_{1}\lambda_{X}+A_{1}{}_{X}\right)+2A_{1}\lambda_{X\text{\ensuremath{\varphi}}}\right] \, , \\
 g_{1} & = & 6\left\{ 2F\lambda_{\varphi}+F_{\varphi}+ 
 X\left[
 Q_{X}+2F_{\text{\ensuremath{\varphi}}X}+4F\lambda_{X\text{\ensuremath{\varphi}}}+
 \lambda_{\varphi}\left(4F_{X}+12F\lambda_{X}- 6A_{1}\right) \right]
 \right.\\
 &&
 \left.
  -4X^{2}\left[\lambda_{\varphi}\left(3A_{1}\lambda_{X}+A_{1}{}_{X}\right)+\A_{1}\lambda_{X\text{\ensuremath{\varphi}}}\right] \right\} \, , \\
g_{2} & = &  6F+6X\left(6F\lambda_{X}+2F_{X}- 3A_{1}\right)-12X^{2}\left(3A_{1}\lambda_{X}+A_{1}{}_{X}\right) \, , \\
g_{3} & = &\Fz-X \left(4F\lambda_{\varphi\varphi}+6F\lambda_{\varphi}^{2}+4F_{\varphi}\lambda_{\varphi}+2F_{\varphi\varphi}+Q_{\varphi}\right) 
+ X{}^{2}\left[(4\lambda_{\varphi\varphi}+6\lambda_{\varphi}^{2})A_1+4\lambda_{\varphi}A_{1\varphi} \right]\, , \\
g_{4} & = & 2(F-A_{1}X) \, ,\\
g_{5} & = & 4\left[F_{\varphi}- X A_{1}{}_{\varphi}+3\lambda_{\varphi}(F-A_{1}X)\right] \, , \\
g_{6} & = & 
2F_{\varphi}+4 \lambda_{\varphi} F
+2X\left[
Q_{X}+2F_{\text{\ensuremath{X\varphi}}}+4\lambda_{X\text{\ensuremath{\varphi}}}F
+ \lambda_{\varphi}(4 F_{X}+12 \lambda_{X} F -6 A_1)
\right] \\
 & &  -8X{}^{2}\left(\lambda_{\varphi}(A_{1}{}_{X}+3\lambda_{X}A_{1})+\lambda_{X\text{\ensuremath{\varphi}}}A_1 \right) \, ,
  \\
g_{7} & = & -8
\left[F_{X}+3\lambda_{X}F- A_1
-X(A_{1X}+3\lambda_{X}A_{1})\right] \, ,
\end{eqnarray*}

The coefficients entering in the scalar equation \eqref{SFE} through the functions \eqref{UandJ} are given by
\begin{eqnarray*}
g_{8} & = & P_\varphi+3\lambda_\varphi P
+3X\left(2F_{\varphi\varphi}\lambda_{\varphi}+4F\lambda_{\varphi}\lambda_{\varphi\varphi}+2F_{\varphi}\lambda_{\varphi\varphi}+6F\lambda_{\varphi}^{3}+8F_{\varphi}\lambda_{\varphi}^{2}+3Q\lambda_{\varphi}^{2}+\lambda_{\varphi}Q_{\varphi}+Q\lambda_{\varphi\varphi}\right)\\
 &  & - 6X^2\lambda_{\varphi}\left(2A_{1}\lambda_{\varphi\varphi}+3A_{1}\lambda_{\varphi}^{2}+A_{1}{}_{\varphi}\lambda_{\varphi}\right) \, , 
 \\
g_{9} & = & -3\left(12F\lambda_{\varphi}^{2}+10F_{\varphi}\lambda_{\varphi}+4F\lambda_{\varphi\varphi}+2F_{\varphi\varphi}+Q_{\varphi}+3\lambda_{\varphi}Q\right)
+12X(3A_{1}\lambda_{\varphi}^{2}+A_{1}{}_{\varphi}\lambda_{\varphi} +A_{1}\lambda_{\varphi\varphi}) \, , 
\\
g_{10}& =&-6\left(F_{\varphi}+3 \lambda_{\varphi}F\right)  +6X\left(A_{1}{}_{\varphi}+3\lambda_{\varphi}A_{1}\right)\, , 
\\
g_{11} & = & 
-\left(Q_{\varphi}+3\lambda_{\varphi}Q\right)
- 6X(6F_{\varphi}\lambda_{\varphi}\lambda_{X}+2F_{\varphi\varphi}\lambda_{X}+2F_{\varphi}\lambda_{X\text{\ensuremath{\varphi}}}+Q_{\varphi}\lambda_{X}+3Q\lambda_{\varphi}\lambda_{X}+Q\lambda_{X\text{\ensuremath{\varphi}}})\, , 
   \\
g_{12} & = & 12  \lambda_{\varphi}(F_{\varphi}+\lambda_{\varphi} F)
+2(P_{X}+3 \lambda_{X} \Fz)
-Q_{\varphi} +3\lambda_{\varphi}Q \\
&& +6X  (-4A_{1}\lambda_{\varphi}^{2}+2F_{X\text{\ensuremath{\varphi}}}\lambda_{\varphi}+2F_{X}\lambda_{\varphi}^{2} +6F\lambda_{\varphi}^{2}\lambda_{X}-2F_{\varphi\varphi}\lambda_{X}+4F\lambda_{\varphi}\lambda_{X\text{\ensuremath{\varphi}}}+\lambda_{\varphi}Q_{X}-Q_{\varphi}\lambda_{X})\\
  &-&12X^{2} (A_{1}{}_{X}\lambda_{\varphi}^{2}+3A_{1}\lambda_{\varphi}^{2}\lambda_{X} +24A_{1}\lambda_{\varphi}\lambda_{X\text{\ensuremath{\varphi}}})\, , 
  \\
g_{13} & = & 12F\lambda_{\varphi}+6F_{\varphi}+
6X\left(-6A_{1}\lambda_{\varphi}+2F_{X\text{\ensuremath{\varphi}}}+4F_{X}\lambda_{\varphi}+12F\lambda_{\varphi}\lambda_{X}+4F\lambda_{X\text{\ensuremath{\varphi}}}+Q_{X}\right)\\
 & &-24X^{2}  \left(A_{1}{}_{X}\lambda_{\varphi}+3A_{1}\lambda_{\varphi}\lambda_{X}+A_{1}\lambda_{X\text{\ensuremath{\varphi}}}\right)\, , 
 \\
g_{14}& = & -12\left[F_{X}+ 3\lambda_{X} F   -A_1-X(A_{1}{}_{X}+3\lambda_{X}A_{1})\right]\, .
\end{eqnarray*}

\section{Effective parameters in the quadratic action of perturbations}
\label{EFTparam}
In this section, we recall the expressions of the effective parameters entering in the quadratic action of the perturbations about
a FLRW background,
 \bea
\begin{split}
& S^\quadac = \int d^3x \,  dt \,  a^3  \frac{M^2}2\bigg\{ \delta K_{ij }\delta K^{ij}- \left(1+\frac23\aL\right)\delta K^2  +(1+\alphaT) \bigg( \R \frac{\delta \sqrt{h}}{a^3} + \delta_2 R \bigg)\\
&  + H^2\alphaK \delta N^2+4 H \alphaB \delta K \delta N+ ({1+\alphaH}) \R  \delta N   +  4 \bun  \delta K  {\delta \dot N }   + \bdeux  {\delta \dot N}^2 +  \frac{\btrois}{a^2}(\partial_i \delta N )^2   
\bigg\} \; ,
\end{split}
\eea
in terms of the functions (evaluated in the background solution) entering in the DHOST action,
\begin{equation}
S=\int d^4x\sqrt{-g}\left(\Fz(X,\varphi)+Q(X,\varphi)\, \Box \varphi+F(X,\varphi)\,R+\sum_{i=1}^{5}A_{i}(X,\varphi)\, L_{i}\right)\, .
\end{equation}
We restrict ourselves to shift-symmetric theories  where all the functions in the action above depend on $X$ only.

All parameters but $\alphaK $ and $\alphaB$ depend on $F$ and $A_i$ only, and they were given in \cite{Langlois:2017mxy},
\beq
\label{alphaandbeta}
\begin{split}
\frac{M^2}{2}=&\, F-A_1 X\,, \qquad \frac{M^2}{2}(1+\a_T)= F\,, \qquad  \frac{M^2}{2}(1+\a_H)=F-2X F_{X} \,,
\\
\frac{M^2}{2}\left(1+\frac23\aL\right)=&\, F+A_2 X\,,\qquad \frac{M^2}{2} \bdeux=-X \left(A_1+A_2+(A_3+A_4) X+A_5X^2\right)\,,
\\
2 M^2\bun=&\, X (4F_{X}+2A_2+A_3X) \,, \quad \frac{M^2}{2} \beta_3=-X(4F_{X}-2A_1-A_4 X)\,,
\end{split}
\eeq
where the right-hand side quantities are evaluated on the homogeneous and isotropic background.

The expressions of $\alphaK $ and $\alphaB$ are much more complicated and they involve, in addition to $F$ and $A_i$, the functions $P$
and $Q$. A long calculation gives,
\bea
2 H M^2 \alphaB & = & (4 H X + \dot X)A_1 + 2(3H X + \dot X) A_2 + \frac{3}{2} X(-2H X + \dot X) A_3 - X\dot X A_4 - X^2 \dot X A_5 \nonumber \\
&+& 4H X^2 A_{1X} + 2X(6HX + \dot X) A_{2X} + X^2 \dot X A_{3X}  \nonumber \\
&+& (-4 H X + 6 \dot X) F_{X} + 2 \sqrt{-X} X Q_X + 4 X \dot X P_{XX} \, , \\
\frac{M^2}{2} H^2 \alphaK & =& (-3H^2 X + 3H \dot X - \frac{3\dot X^2}{2X} + 2 \ddot X) A_1 + (9 \dot H X + 3H \dot X -  \frac{3\dot X^2}{2X} + 2 \ddot X) A_2 \nonumber \\
&+& \frac{3}{4} (18 H^2 X^2 + 10 \dot H X^2 + 8 H X \dot X - \dot X^2 + 4 X \ddot X) A_3 + (6H X \dot X - \frac{3X^2}{4} + 3X \ddot X) \nonumber \\
&+& X(9H X \dot X + \dot X^2 + 4X \ddot X) A_5 + (-15 H^2 X^2 + 3HX\dot X + \frac{3 \dot X^2}{4} + X \ddot X) A_{1X}  \nonumber \\
&+& (-27 H^2 X^2 + 3H X \dot X + \frac{3 \dot X^2}{4} + 6 X^2 \dot H  + X \ddot X) A_{2X} 
+\frac{X}{4} (12 H X \dot X + 7 \dot X^2 + 4 X \ddot X)A_{4X}
\nonumber \\
&+& (9H^2 X^3 + 3 \dot H X^3 + 3H X^2 \dot X + \frac{7 X \dot X^2}{4} +X^2 \ddot X) A_{3X} 
 + \frac{X^2}{4}(12 H X \dot X + 11 \dot X^2 + 4 X \ddot X) A_{5X} \nonumber \\
 &+& (-6H^2 X^3 + \frac{X \dot X^2}{2}) A_{1XX} + (-18 H^2 X^3 + \frac{X \dot X^2}{2}) A_{2XX} + \frac{X^2 \dot X^2}{2} A_{3XX} \nonumber \\
 &+& \frac{X^2 \dot X^2}{2} A_{4XX} + \frac{X^3 \dot X^2}{2} A_{5XX} + 6X(2H^2 + 3 \dot H)  F_{X}  \nonumber \\
 &+& 12 X^2 (2 H^2 + \dot H) F_{XX} + 2 X^2 Q_{XX} - 6H X^2 \sqrt{-X} Q_{XX} \, .
\eea
When applied to the model we are considering in the paper, 
\begin{eqnarray}
&&\Fz  =  \cP X, \quad Q  =  0,\quad F  =  \frac{1}{2},\quad A_{2}  =  \cA X,\quad A_{3}=-2(\cA+2\cL)-8 \cA \cL X^{2} \, , \nonumber \\
&&A_4 =  2(\beta + 2\mu- 2 \mu^2 X^2) \, , \quad A_5 = 8\mu X (\beta + 2 \mu + 3 \beta \mu X^2) \, , \label{A4A5}
\end{eqnarray}
the expressions of \eqref{alphaandbeta}  yield
\beq
\begin{split}
{M^2}=&\, 1+ 2 \beta X\,, \qquad \a_T= \a_H = - \frac{2 \beta X^2}{1+2\beta X^2}\,,  \qquad \alpha_L = 0 \, , 
\\
\bun=&\, -2 \mu X^2 \,, \qquad \beta_2 = -24 \mu^2 X^4 \, , \qquad \beta_3 = - \frac{8\mu X^2(-1+ \mu X^2)}{1+ 2 \beta X^2} \, ,
\end{split}
\eeq
while the expressions for $\alphaB$ and $\alphaK$ simplify into
\bea
\alpha_{B} & = & \frac{2 \dot{\varphi }^3 \left(4 \cA \dot{\varphi } H+\cL \left(6 \cA X^2 \left(\dot{\varphi } H-3 \dot{\varphi}\right)+3 \dot{\varphi } H-5\ddot{ \varphi}\right)-6 \cL^2 X^2 \left(2 \cA X^2+1\right)\ddot{ \varphi }\right)}{2 \cA X^2 H+H},\\
\alpha_{K} & = & -\frac{2X}{H^2 \left(2 \cA X^2+1\right)} (6 X (2 \cL^2 X (24 \dot{\varphi } H (3 \cA X^2+1) \ddot{\varphi }+2 \dot\varphi  \dddot \varphi (14 \cA X^2+5) \nonumber\\
&+& (126 \cA X^2+25) (\ddot{\varphi })^2)+\cL (6 \cA X^2 (7 H^2+3 \dot{H})+9 H^2+5 \dot{H})+6 \cA H^2)-\cP ) ,
\eea
where we have used $X=-\dot \varphi^2$, $\dot X=-2 \dot\varphi \ddot\varphi $ and $\ddot X=-2 (\ddot\varphi^2 + \dot\varphi \dddot \varphi)$. 

In terms of the variables introduced in (\ref{defofparam}), these coefficients become
\bea
M^{2} & = & 1+2x_{2},\quad \alpha_{T}=\alpha_{H}=-\frac{2x_{2}}{2x_{2}+1},\quad \alpha_{L}=0\,, \nonumber \\
\beta_{1} &= & -2sx_{2}\,, \quad \beta_{2}=-24s^{2}x_{2}^{2} \, , \quad \beta_{3}  =-\frac{8sx_{2}\left(sx_{2}-1\right)}{2x_{2}+1}\, , \nonumber \\
\alpha_{B} & = & -\frac{2 x_2 \left(5 \left(2 s x_2+s\right) \epsilon _{\varphi }-6 s x_2-3 s-4\right)}{\left(2 x_2+1\right) \left(2 s x_2 \epsilon _{\varphi }+1\right)} \, , \nonumber \\
\alpha_{K}& = &\frac{2}{(2 x_2+1) (2 s x_2 \epsilon _{\varphi }+1)^2}\left(x_1+\epsilon _{\varphi } \left(x_2^3 \left(336 s^2 \epsilon _h-144 s^2\right)+24 s x_2^2 \left(5 s \epsilon _h+3 s-6\right)\right)\right. \nonumber\\
 & - & \left. (\dot{H}/H^2) \left(\left(432 s^3 x_2^4+120 s^3 x_2^3\right) \epsilon _{\varphi }^2+\left(432 s^2 x_2^3+120 s^2 x_2^2\right) \epsilon _{\varphi }+108 s x_2^2+30 s x_2\right)\right. \nonumber\\
 & + & \left. \epsilon _{\varphi }' \left(\left(672 s^3 x_2^4+240 s^3 x_2^3\right) \epsilon _{\varphi }+336 s^2 x_2^3+120 s^2 x_2^2\right)\right. \nonumber\\
 & + & \left.\left(720 s^3 x_2^4+24 s^2 (15 s+71) x_2^3+420 s^2 x_2^2\right) \epsilon _{\varphi }^2-252 s x_2^2-(54 s+36) x_2\right)\, .
\eea
Notice that in the last equation, we have used the relation $\dddot \varphi =\epsilon_{\varphi}'(1+2 s x_2\epsilon_{\varphi})+\epsilon_{\varphi}^{2}+\epsilon_{\varphi}\epsilon_{h}$, which can be deduced from the relations 
\beq
\dot X=-2 \dot\varphi \ddot\varphi =2 \epsilon_\varphi H_b X\,,
\eeq
which comes from the definition of $\epsilon_\varphi$,
and
\beq 
\ddot X=-2 (\ddot\varphi^2 + \dot\varphi \dddot \varphi)= 2 X H_b^2\left(\frac{H}{H_b}\epsilon_\varphi^\prime +\epsilon_\varphi \epsilon_h+2\epsilon_\varphi^2\right)\,.
\eeq
We could also replace $\dot H/H^2$ in the expression for $\alpha_K$  by using the relation 
\begin{equation}
\label{HdtH2}
\frac{\dot{H}}{H^2}=\frac{\epsilon_h + 2 s (x_2' \epsilon_\varphi + x_2 \epsilon_\varphi')}{1+2 s x_2 \epsilon_\varphi},
\end{equation}
which follows from (\ref{H_Hb}).

Finally, these results allow us to express the coefficients $A_\zeta$ in \eqref{Azeta} and $B_\zeta$ in \eqref{Bzeta}, entering in the 
quadratic action for the scalar perturbation, in terms of the dynamical variables in the form,
\bea
A_\zeta & = & \frac{{\cal A}_{1}(x_2) + x_1 {\cal A}_{2}(x_2)}{ {\cal A}_{3}(x_2)} \, , \\
B_\zeta & = & \frac{{\cal B}_{1}(x_2) + x_1 {\cal B}_{2}(x_2) + x_1^2 {\cal B}_{3}(x_2)}{{\cal B}_{4}(x_2) + x_1 {\cal B}_{5}(x_2)} \, ,
\eea
where the functions ${\cal A}_{i}$ and ${\cal B}_{i}$ are polynomials of the variable $x_2$ only (which depends on the parameter $s$) given by
\begin{eqnarray*}
{\cal A}_{1}(x_2) &= & 6 x_2 \left(216 s^2 x_2^3+\left(54 s^2+84 s+40\right) x_2+36 s (6 s+5) x_2^2-3 (s+4) \right)\, , \\
{\cal A}_{2}(x_2)  & = & 2 \left(2 x_2+1\right) \left(-18 s^2 x_2^2+15 s x_2+1\right) \, , \\
{\cal A}_{3}(x_2)  & = & \left(12 s x_2^2+2 (3 s+5) x_2+1\right)^2 \, , \\
{\cal B}_{1}(x_2)  & = & 3 x_2 \left(38016 s^4 x_2^7 +576 s^3 (132 s+125) x_2^6+288 s^2 \left(198 s^2+344 s+215\right) x_2^5\right. \nonumber\\
 & +&   16 s \left(1188 s^3+2538 s^2+2784 s+1775\right) x_2^4-4 \left(279 s^3+1176 s^2+1154 s+500\right) x_2^2 \nonumber\\
 & + & \left. \left(120+96 s-90 s^2\right) x_2+8 \left(297 s^4+288 s^3-282 s^2+600 s+500\right) x_2^3+9 (3 s+4)\right) \, , \\
{\cal B}_{2}(x_2)  & = & \, -3456 s^4 x_2^7-576 s^3 (9 s-20) x_2^6-96 s^2 
\left(27 s^2-240 s - 76  \right) x_2^5
\nonumber\\
 &-&16 s \left(27 s^3-900 s^2-822 s-280\right) x_2^4 
  +  40 \left(72 s^3+96 s^2+32 s+47\right) x_2^3
 \nonumber\\
 & -& 4 \left(114 s^2+400 s+21\right) x_2^2
 -2 (88 s+67) x_2-3 \, , \\
{\cal B}_{3}(x_2)  & = & \left(2 x_2+1\right) \left(432 s^4 x_2^5+216 s^3 (s+3) x_2^4+12 s^2 (15 s+1) x_2^3 \right. \nonumber \\
&- & \left. 2 s (45 s+101) x_2^2  - 5 (5 s+2) x_2-1\right)\,, \\
{\cal B}_{4}(x_2)  & = &  3 x_2\left(2 x_2+1\right)\left(12 s x_2^2+2 (3 s+5) x_2+1\right){}^2 \left[18s^2 x_2 \left(2 x_2+1\right)^2+60s x_2^2+4(3s+10)x_2-3(3s+4)\right]\,, \\
{\cal B}_{5}(x_2)  & = & \left(2 x_2+1\right)^2 \left(12 s x_2^2+2 (3 s+5) x_2+1\right){}^2 \left(18 s^2 x_2^2 +21 s x_2+1\right) \, .
\end{eqnarray*}
These are the expressions we use to plot Fig. \eqref{plotw2} and to express the eigenvalues \eqref{eigeninmatter} in the matter era 
where $|x_1| \ll|x_2| \ll 1$.

\bibliographystyle{utphys}
\bibliography{Dark_Energy_biblio_20}
\end{document}